\documentclass[showpacs,showkeys,amsmath,amssymb,superscriptaddress,reprint, preprintnumbers,nofootinbib,notitlepage,aps,prl,twocolumn]{revtex4-1}

\pdfoutput=1 
\usepackage[normalem]{ulem}
\usepackage{amsmath}
\usepackage{amssymb}
\usepackage[font=small]{caption}
\usepackage[utf8]{inputenc}
\usepackage[english]{babel}
\usepackage{bm}
\usepackage{graphicx}
\usepackage{dcolumn}
\usepackage{pbox}
\usepackage{epsfig}
\usepackage[dvipsnames]{xcolor}
\usepackage{slashed}
\usepackage{amssymb}
\usepackage{textcomp}
\usepackage{gensymb}
\usepackage{ mathrsfs }
\usepackage{color}
\usepackage[font=small]{caption}
\usepackage[font=small]{subcaption}
\usepackage{url}
\usepackage{comment}
\definecolor{DarkBlue}{rgb}{0.7, 0.4, 1} 
\definecolor{Blue}{rgb}{0, 0.8, 0} 
\definecolor{MyLightBlue}{rgb}{0.5,0.7,1.9}
\definecolor{MyGreen}{rgb}{0.0,0.2, 0.0}
\definecolor{MyBrickRed}{rgb}{0, 0.5, 0.2}
\RequirePackage{hyperref}
\hypersetup{colorlinks, citecolor=Blue,linkcolor=DarkBlue, urlcolor=Green}
\usepackage[normalem]{ulem}

\newcommand{\bea}{\begin{eqnarray}}
\newcommand{\eea}{\end{eqnarray}}

\makeatletter

\renewcommand\@makecaption[2]{%
  \par
  \vskip\abovecaptionskip
  \begingroup
  
   \small\rmfamily
    \begingroup
     \samepage
     \flushing
     \let\footnote\@footnotemark@gobble
     \@make@capt@title{#1}{#2}\par
    \endgroup
  \endgroup
  \vskip\belowcaptionskip
}
\makeatother

\DeclareUnicodeCharacter{2212}{-}
\newcommand{\mdm}{m_{\rm dm}}
\newcommand{\Trh}{T_\text{rh}}

\newcommand{\gs}{g_\star}
\newcommand{\gss}{g_{\star s}}
\begin{document}
\title{Constraining Gravitational Dark Matter with LHAASO and Fermi-LAT}
\author{Basabendu Barman}
\email{basabendu.b@srmap.edu.in}
\affiliation{\,\,Department of Physics, School of Engineering and Sciences, SRM University-AP, Amaravati 522240, India}
\author{Arindam Das}
\email{arindamdas@oia.hokudai.ac.jp}
\affiliation{\,\,Institute for the Advancement of Higher Education, Hokkaido University, Sapporo 060-0817, Japan}
\affiliation{\,\,Department of Physics, Hokkaido University, Sapporo 060-0810, Japan}
\author{Prantik Sarmah}
\email{prantiksarmah@ihep.ac.cn}
\affiliation{\,\,State Key Laboratory of Particle Astrophysics, Institute of High Energy Physics, Chinese Academy of Sciences, Beijing, 100049, China}
\author{Rakesh Kumar SivaKumar}
\email{rakesh_sivakumar@srmap.edu.in}
\affiliation{\,\,Department of Physics, School of Engineering and Sciences, SRM University-AP, Amaravati 522240, India}
\begin{abstract}   
We use diffuse Galactic high energy gamma ray data from LHAASO and Fermi-LAT to constrain gravitationally produced decaying dark matter (DM). Focusing on four benchmark candidates: a dark photon, a heavy right-handed neutrino (RHN), a pseudo–Nambu–Goldstone boson (pNGB), and a non-minimally coupled scalar we derive bounds on the DM mass and its couplings to the visible sector. For dark photons, RHNs, and pNGBs, the combined data constrain the effective interaction strength to $\lesssim\mathcal{O}(10^{-30})$ for DM masses $\gtrsim\mathcal{O}$(TeV). For the non-minimally coupled scalar, this interaction strength is limited to $\lesssim\mathcal{O}(10^{-10})$, for the same DM mass range. Moreover, photon–dark photon oscillations yield strong constraints for massive dark photon beyond 10 GeV, closing a region of parameter space previously left unconstrained.
\end{abstract}
\maketitle
\noindent
{\textbf{Introduction}.--}
Among the four fundamental forces, gravity is by far the weakest. Yet it possesses a unique universality: it couples indiscriminately to all forms of energy and matter known to exist. This makes gravity the most natural mediator between dark matter (DM), which accounts for roughly 24\% of the total matter–energy content of the Universe~\cite{Planck:2018vyg}, and the visible sector described by the Standard Model (SM). Although gravitational interactions are Planck-suppressed, DM can still be efficiently produced at very high temperatures, most notably during reheating, realizing the UV freeze-in mechanism~\cite{Elahi:2014fsa,Bernal:2017kxu}. If no exact symmetry enforces cosmological stability, such DM may decay into SM final states demanded by the underlying particle physics model, which may also address open questions such as neutrino mass generation. A particularly intriguing aspect, relevant for the present work, is that the DM mass range naturally associated with gravitational production may contribute to the diffuse Galactic gamma ($\gamma$)-ray flux observed by telescopes such as LHAASO~\cite{LHAASO:2023gne,LHAASO:2024lnz} and Fermi-LAT~\cite{Zhang:2023ajh,Fermi-LAT:2009ihh}. For high-energy diffuse Galactic $\gamma$-ray observations, LHAASO, through its SKA and WCDA arrays, has provided unprecedented measurements from the TeV to PeV range~\cite{LHAASO:2024lnz}, while Fermi-LAT has mapped the corresponding GeV to TeV emission~\cite{Zhang:2023ajh}, covering both the inner ($15^\circ < l < 125^\circ$, $|b| < 5^\circ$) and outer ($125^\circ < l < 235^\circ$, $|b| < 5^\circ$) Galactic plane regions~\cite{LHAASO:2023gne,Zhang:2023ajh}.

In this {\it letter}, we explore several well-motivated scenarios, rooted in particle physics, gravity, and cosmology, in which DM is produced through minimal or non-minimal gravitational interactions in the light of observed high energy Galactic $\gamma$-rays by LHAASO~\cite{LHAASO:2023gne} and Fermi-LAT~\cite{Zhang:2023ajh}. 
Specifically, we have `four horsemen' of DM candidates: the dark photon, the right-handed neutrino (RHN), a massive pseudo–Nambu–Goldstone boson (pNGB), and a non-minimally coupled scalar singlet, respectively. Following their gravitational production, these DM species can decay into visible particles, with the decay channels determined by the structure of their interactions with the visible sector. The decay of these DM candidates into photons contributes to the Galactic diffuse $\gamma$-ray emission. Using LHAASO and Fermi-LAT data, we constrain the resulting $\gamma$-ray fluxes, which in turn provides bound on the corresponding DM decay rates or their interaction strengths with the SM. Remarkably, the combined constraints from LHAASO and Fermi-LAT probe regions of parameter space characterized by extremely feeble couplings  $\lesssim\mathcal{O}(10^{-30})$, depending on the underlying particle physics model and very heavy DM~\cite{Boehm:2025qro,Dubey:2025ouh}, domains those have hitherto remained largely unexplored within these theoretical frameworks. We also derive limits on photon–dark photon oscillations induced by kinetic mixing, that surpass existing bounds for dark photon masses above $\sim 10$~GeV.
\\

\begin{figure*}
    \centering    \includegraphics[width=0.47\linewidth]{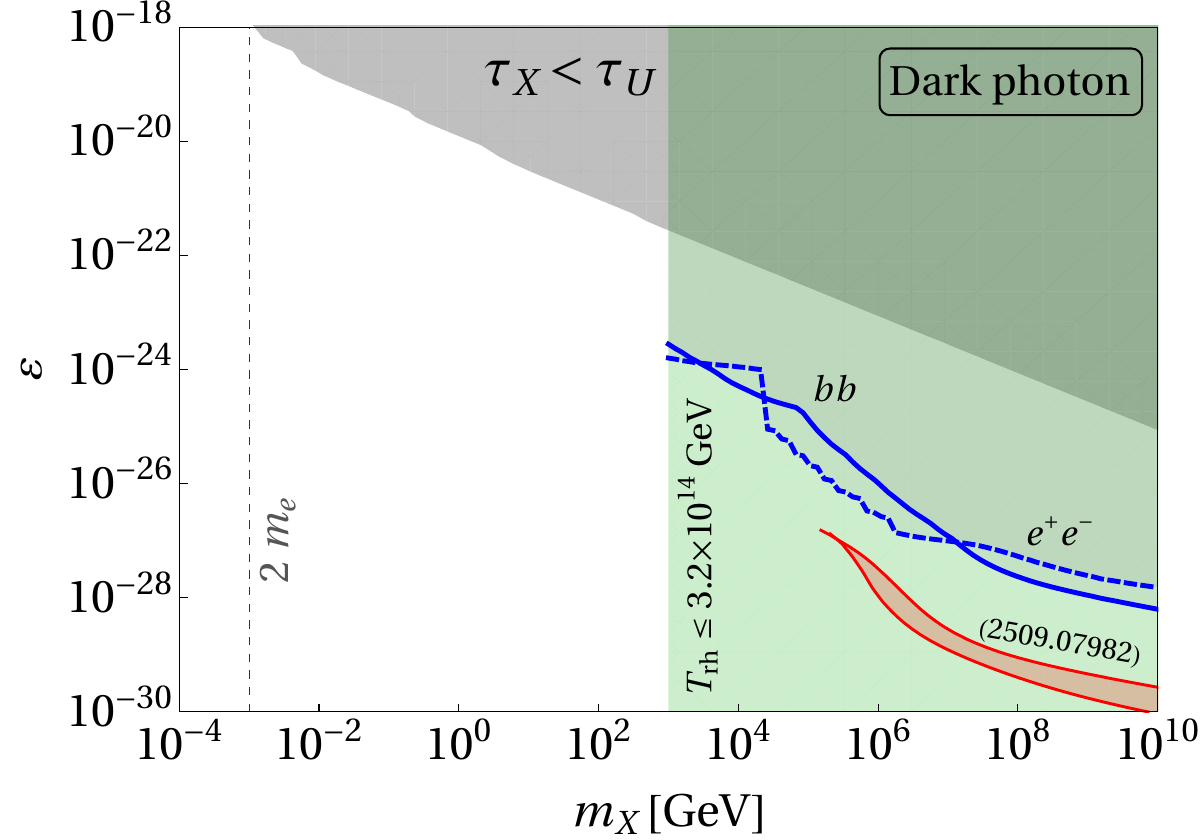}~ \includegraphics[width=0.47\linewidth]{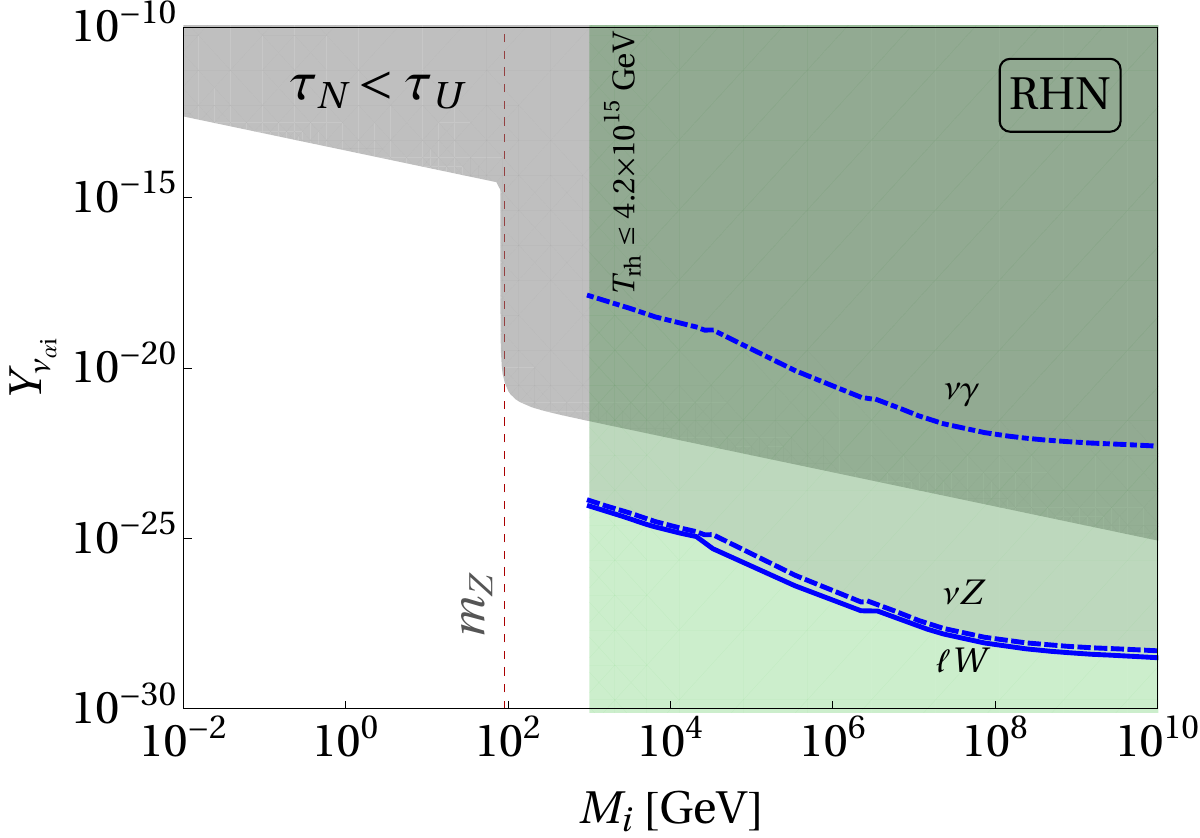}\\[10pt] \includegraphics[width=0.47\linewidth]{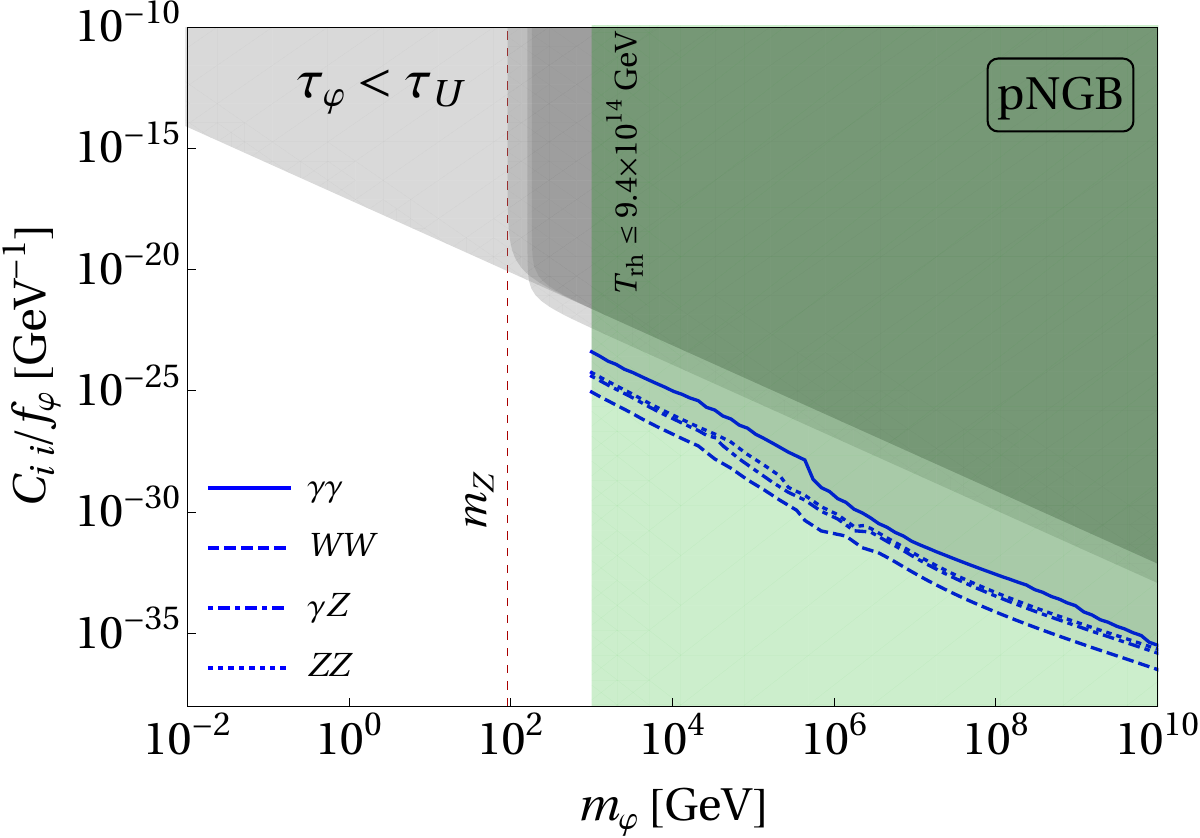}~\includegraphics[width=0.47\linewidth]{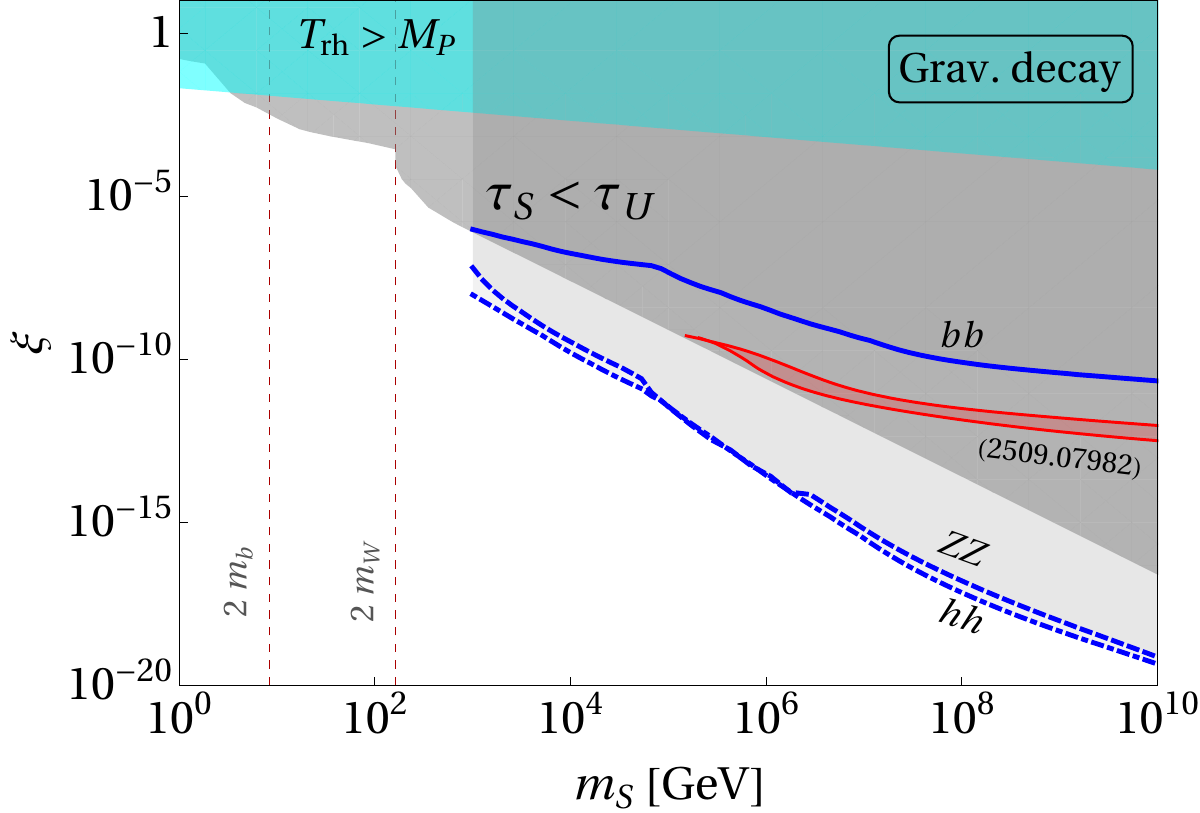}
    \caption{Bounds on DM coupling and mass, depending on different benchmark scenarios. Within the gray shaded region the lifetime of the relevant DM species is below the lifetime of the Universe. The blue curves are obtained utilizing combined data from LHAASO and Fermi-LAT. Within the green shaded region right relic abundance is obtained. The cyan shaded region in the bottom right panel is forbidden from $\Trh>M_P$ [cf.Eq.~\eqref{SDM1}]. {\color{black} The red filled bands in the top left and bottom right panels are bounds extracted from Fig.~4 of Ref.~\cite{Boehm:2025qro}, corresponding to 2-body decay of DM into $b\bar{b}$ final states (see text for details)}.}
    \label{fig:plt1}
\end{figure*}
\noindent
{\textbf{Frameworks.}--}
We now turn to scenarios in which particle DM is produced gravitationally—either through massless-graviton mediation or via a non-minimal gravity portal. Once produced, such DM can subsequently decay into photons, either directly or through intermediate states, depending on the 
underlying particle-physics model manifesting the DM metastable. This leads to a key constraint on its lifetime, namely $\tau_{\rm DM} > \tau_U \simeq 10^{17}~\text{s}$, where $\tau_U$ denotes the age of the Universe. In the following, we discuss each scenario in detail.
\\

$(i)$ {\it Gravity portal to decaying DM:} Minimal gravity-mediated production of DM from the radiation bath in the early Universe is an unavoidable irreducible process. Such gravitational production of frozen in DM has been studied in, e.g., Refs.~\cite{Ema:2015dka,Garny:2015sjg,Tang:2016vch,Ema:2016hlw,Garny:2017kha,Tang:2017hvq,Bernal:2018qlk,Clery:2021bwz}. The relevant Lagrangian in this case,
\begin{align}
\mathcal{L}\supset-\frac{1}{M_P}\,h_{\mu\nu}\,T^{\mu\nu}\,,    
\end{align}
where $h_{\mu\nu}$ is the massless spin-2 graviton and $T^{\mu\nu}$ is the energy-momentum tensor, that depends on the spin of the DM. The interaction rate density for a 2-to-2 DM production process reads $\gamma(T)=k\,T^8/M_P^4$~\cite{Garny:2015sjg,
Barman:2021ugy},
where $k\simeq 2.9\times 10^{-3}$ (spin-0), $k\simeq 3.4\times 10^{-2}$ (spin-1/2 Majorana) or  $k\simeq 7.3\times 10^{-2}$ (spin-1), considering $T\gg\mdm$ \cite{
Clery:2022wib,Barman:2022qgt,Garcia:2023obw}. In order to follow the evolution of the DM number density $n_{\rm DM}$, we express the Boltzmann equation in terms of the DM yield {\color{black}$Y_{\rm DM}\equiv n_{\rm DM}/\mathfrak{s}$} as,  
\begin{equation}\label{eq:beq}
x\,\mathcal{H}\,\mathfrak{s}\,\frac{dY_{\rm DM}}{dx} =\gamma(T)\,,
\end{equation}
where $x \equiv m_{\rm DM}/T$ is a dimensionless variable with $T$ the bath temperature. For a radiation-dominated Universe, the entropy density $\mathfrak{s}(T)$ and the Hubble rate $H(T)$ are given by  
\begin{align}
&\mathfrak{s}(T)=\frac{2\pi^2}{45}\,\gss(T)\,T^3,\,\, \,\, \, \, 
\mathcal{H}(T)=\frac{\pi}{3}\,\sqrt{\frac{\gs(T)}{10}}\,\frac{T^2}{M_P}\, ,
\end{align}
where $\gss(T)$ and $\gs(T)$ denote the effective relativistic degrees of freedom associated with entropy and energy densities, respectively and $M_P$ is the reduced Planck mass. To account for the observed relic abundance, the present-day DM yield must satisfy  $Y_0\,m_{\rm DM} = \Omega h^2 \,\frac{1}{{\color{black} \mathfrak{s}_0}}\,\frac{\rho_c}{h^2}\simeq  4.3\times 10^{-10}\,{\rm GeV}$
where $Y_0 \equiv Y_{\rm DM}(T_0)$ is DM yield at present epoch, $\rho_c \simeq 1.05 \times 10^{-5} h^2\,{\rm GeV/cm}^3$ is the critical energy density, ${\color{black} \mathfrak{s}_0}\simeq 2.69 \times 10^3\,{\rm cm}^{-3}$ the present entropy density~\cite{Planck:2018vyg} and $\Omega h^2 \simeq 0.12$ the measured DM relic density~\cite{Planck:2018vyg}. Solving Eq.~\eqref{eq:beq}, one can analytically obtain the asymptotic yield,
\begin{align}\label{eq:Y0}
&Y_0\simeq\frac{45\, k}{2\pi^3\, \gss} \sqrt{\frac{10}{\gs}} \left(\frac{\Trh}{M_P}\right)^3\,, 
\end{align}
leading to reheating temperature
\begin{equation}
\Trh\simeq\left(\frac{1\,\text{PeV}}{\mdm}\right)^{1/3}
\begin{cases}
9.4\times10^{14}\,\text{GeV} & \text{spin-0}\,,
\\[10pt]
5.3\times10^{14}\,\text{GeV} & \text{spin-1/2}\,,
\\[10pt]
3.2\times10^{14}\,\text{GeV} & \text{spin-1}\,,
\end{cases}
\end{equation}
in order to satisfy the right DM abundance. These DM, once produced, could undergo decay into photonic final states, depending on the following well-known benchmark scenarios:
\begin{itemize}
\item [$\color{black}\bullet$] {\it Decaying dark photon via kinetic mixing:} An extended gauge sector containing two $U(1)$ symmetries, one associated with the visible sector and the other with a dark sector composed of SM gauge singlets. These two sectors can interact via a kinetic mixing portal~\cite{Holdom:1985ag}, which gives rise to the interaction Lagrangian
\begin{equation}\label{eq:kinmix}
\mathcal{L}\supset -\frac{1}{4}\,X^{\mu\nu}\,X_{\mu\nu}-\frac{\varepsilon}{2}\,F^{\mu\nu}\,X_{\mu\nu}-\frac{1}{2}\,m_X^2\,X_\mu\,X^\mu\,,
\end{equation}
where, $F^{\mu\nu}=\partial^{\mu}A^{\nu}-\partial^{\nu}A^{\mu}$ and $X^{\mu\nu}=\partial^{\mu}X^{\nu}-\partial^{\nu}X^{\mu}$ are the field-strength tensors of the SM and dark $U(1)$ gauge fields, $A^\mu$ and $X^\mu$, respectively. Diagonalizing Eq.~\eqref{eq:kinmix} one arrives at an interaction Lagrangian,
\begin{align}
& \mathcal{L}\supset-\frac{e\,\varepsilon}{\sqrt{1-\varepsilon^2}}\,J_\mu\,X^\mu+e\,J_\mu\,A^\mu\,.    
\end{align}
The dark photon then can decay into the visible sector with decay rates~\cite{Fabbrichesi:2020wbt}
\begin{align}
&\Gamma_X=
\begin{cases}
\displaystyle
\frac{N_c\,Q_f^2\,\alpha_{\rm EM}\,\varepsilon^2}{3}\,m_{X}\,\sqrt{1-\frac{4m_\ell^2}{m_{X}^2}}\,\left(1+\frac{2m_\ell^2}{m_{X}^2}\right)\,, ~f\bar{f}
\\[10pt]
\displaystyle
\frac{\mathbb{R}\alpha_{\rm EM}\,\varepsilon^2}{3}\,m_{X}\,\sqrt{1-\frac{4m_\mu^2}{m_{X}^2}}\,\left(1+\frac{2m_\mu^2}{m_{X}^2}\right)\,,~\text{hadron}\,,
\end{cases}
\end{align}
where $N_c=1(3)$ for leptons (quarks), $Q_f$ is the corresponding fermionic electromagnetic (EM) charge and $\mathbb{R}\equiv\frac{\sigma(e^+e^-\to\text{had})}{\sigma(e^+e^-\to\mu^+\mu^-)}$ is the EM spectral function ratio. 
\item [$\color{black}\bullet$] {\it Photon-dark photon oscillation:} Dark photons can oscillate into SM photons through kinetic mixing, presented in Eq.~\eqref{eq:kinmix}\footnote{In~\cite{Ramazanov:2023nxz}, LHAASO bounds has been utilized to constrain graviton energy via graviton-to-photon conversion.}. 
\begin{figure*}
\centering    \includegraphics[width=0.64\linewidth]{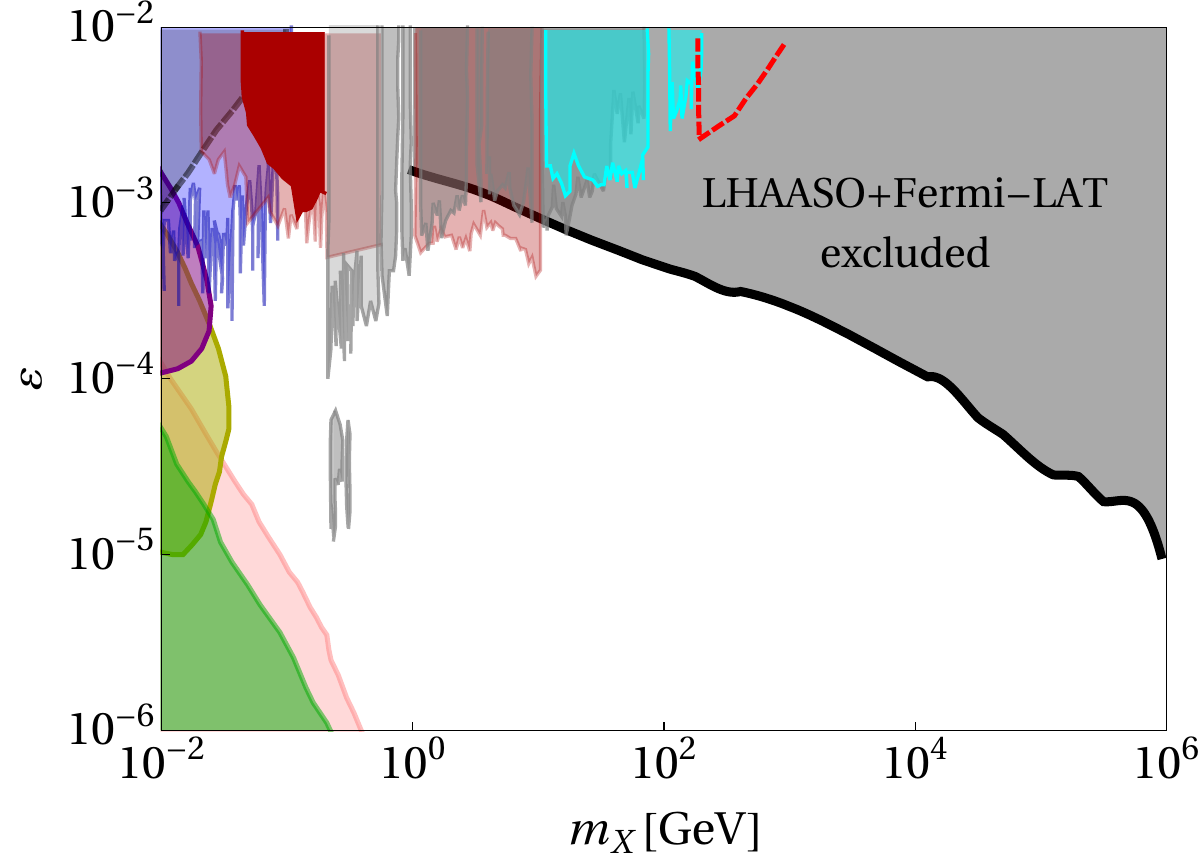}
\caption{Bounds on kinetic mixing parameter $(\varepsilon)$ with respect to dark photon mass $(m_X)$ from oscillation conversion. Combined bounds from LHAASO and Fermi-LAT are shown by the thick black solid line. We show existing limits from  di-lepton searches  at low energy scattering, high energy collider and fixed target experiments: A1~\cite{Merkel:2014avp} (darker red),  LHCb~\cite{Aaij:2019bvg} (gray), 
CMS~\cite{CMS:2019kiy} (cyan),
BaBar~\cite{Lees:2014xha} (lighter red), 
NA48/2~\cite{Batley:2015lha} (blue), as well as beam–dump experiments:  E141~\cite{Riordan:1987aw} (yellow), $\nu$-Cal~\cite{Blumlein:2011mv,Blumlein:2013cua} (pink), CHARM~\cite{Gninenko:2012eq} (green), and projected limit from HL-LHC~\cite{Curtin:2014cca} (red, dashed). Additional limits arises from the electron anomalous magnetic moment $(g-2)_e$~\cite{Pospelov:2008zw} (gray diagonal dashed line).}
    \label{fig:osc}
\end{figure*}
A fraction of dark photon can get converted into photons with flux,
\begin{align}
&\Phi_\gamma(\omega)=\Phi_X(\omega)\cdot P_{X\to\gamma}\,,
\label{eq:oscillation}
\end{align}
where the conversion probability $P_{X\to\gamma}\propto\varepsilon^2$ (see Sec.~\ref{sec:osc} for details), for GeV-PeV scale dark photons and $\Phi_{X}$ is the integrated dark photon flux. 
\item [$\color{black}\bullet$] {\it Decaying Majorana DM:} The minimal $U(1)_X$ extension framework, that addresses both neutrino masses and DM. The three SM gauge singlet right-handed neutrinos (RHN), responsible for light neutrino masses via the seesaw mechanism after $U(1)_X$ and electroweak symmetry breaking (EWSB), also ensures complete cancellation of the gauge anomaly. Details of the anomaly-free $U(1)_X$ construction and neutrino-mass generation can be found in~\cite{Barman:2025bir,Barman:2025hoz}. Partial decay widths of the heavy RHNs are given by,
\begin{align}
&\Gamma_{N_\beta}\simeq
\begin{cases}
\displaystyle
\frac{\lvert Y_{\nu_{\alpha \beta}}^{\rm NH(IH)}\rvert^2}{32 \pi}\,M_{N_\beta}\,,& W\ell_\alpha
\\[10pt]
\displaystyle
\frac{\lvert Y_{\nu_{\alpha \beta}}^{\rm NH(IH)}\rvert^2}{64 \pi}\,M_{N_\beta}\,, & Z\nu_\alpha\,,
\end{cases}
\end{align}
while its radiative decay rate for $M_N \gg m_{W}$ reads, 
\begin{align}
&\Gamma_{N_\beta\rightarrow \nu_\alpha\,\gamma} \simeq M_{N_\beta}\,\frac{\alpha_{\rm EM}\,\lvert Y_{\nu_{\alpha \beta}}^{\rm NH(IH)}\rvert^2}{256\,\pi^6}\,\left(\frac{m_\mu}{v}\right)^2\,,
\end{align}
where $m_{W(\mu)}$ is the $W$ boson (muon) mass in SM. Depending on the normal hierarchy (NH) or inverted hierarchy (IH) scenario, it is possible to consider $N_1$ or $N_3$ to be the viable DM candidate.
\end{itemize}
(ii) {\it Massive pNGB DM:}
We consider an effective field theory describing the interactions of a gauge-singlet pseudoscalar resonance $\varphi$ with the electroweak (EW) gauge bosons of the SM. Using the interactions described in Sec.~\ref{sec:pngb-int}, the partial decay rates are computed as,
\begin{align}\label{eq:pngb-decay}
&\Gamma_\varphi=
\begin{cases}
\displaystyle
\frac{1}{2} \left(\frac{C_{\gamma\gamma}}{f_\varphi}\right)^2
\frac{\alpha^2 m_\varphi^3}{32\pi^3}\,, & \gamma\gamma
\\[10pt]
\displaystyle
\left(\frac{C_{\gamma Z}}{f_\varphi}\right)^2
\frac{\alpha^2 m_\varphi^3}{32\pi^3}
\left[1 - \left(\frac{m_Z}{m_\varphi}\right)^2\right]^3\,, & Z\gamma
\\[10pt]
\displaystyle
\frac{1}{2} \left(\frac{C_{ZZ}}{f_\varphi}\right)^2
\frac{\alpha^2 m_\varphi^3}{32\pi^3}
\left[1 - \left(\frac{2m_Z}{m_\varphi}\right)^2\right]^{3/2}\,, & ZZ
\\[10pt]
\displaystyle
\left(\frac{C_{WW}}{f_\varphi}\right)^2
\frac{\alpha^2 m_\varphi^3}{32 s_W^4 \pi^3}
\left[1 - \left(\frac{2m_W}{m_\varphi}\right)^2\right]^{3/2}\,, & WW
\end{cases}
\end{align}
where $m_{Z,W}$ denotes the $Z(W)$ boson mass. 
\\

(iii) {\it Gravity induced DM decay:} We consider a SM gauge singlet scalar DM $S$ non-minimally coupled to the Ricci scalar $\mathcal{R}$ via a linear coupling of the form $\xi\,M_P\,S\,\mathcal{R}$, where $\xi$ is the strength of the non-minimal interaction. The relevant action in the Jordan frame in this case reads,
\begin{align}\label{eq:nonmin}
& \mathcal{S }= \int d^4x\, \sqrt{-g}\, \Bigg[-\frac14\, g^{\mu\nu}\, g^{\lambda\rho}\, \mathcal{V}_{\mu\lambda}^{(a)}\, \mathcal{V}^{(a)}_{\nu\rho} + 
\nonumber\\&
\frac{1}{\omega^2} \left(\mathcal{L}_Y - V(H)\right) 
+ \frac{1}{\omega}\, \big|D_\mu H\big|^2 + \frac{i}{\omega^{3/2}}\, \bar{f}\, \slashed{\partial}\, f 
\nonumber\\&
+ \frac{3\,i}{\omega^2}\, \bar{f} \left(\slashed{\partial}\, \omega\right) f\Bigg]\,,
\end{align}
using the conformal transformation $\widetilde{g}_{\mu\nu}=\omega^{-1}\,g_{\mu\nu}$ where 
$\omega\simeq1+(\xi/M_P)\,\widetilde{S}$, with $\widetilde{S}=S\,\left(1+\frac{\xi}{4}\,\frac{S}{M_P}\right)$. Clearly, the DM $S$ can decay into all possible SM final states following Eq.~\eqref{eq:nonmin}. These decay channels are purely gravity induced for $\xi\to0$ ensuring the absolute stability of the DM. Hence we obtain the following Planck-suppressed two-body decay rates for the DM,
\begin{align}
& \Gamma_S = \frac{\xi^2\,m_S^3}{32\pi\,M_P^2}
\begin{cases}
\displaystyle
\left[1 - 4\!\left(\frac{m_h}{m_S}\right)^2\right]^{1/2}, 
& hh \\[10pt]
\displaystyle
N_c\!\left(\frac{m_f}{m_S}\right)^2
\left[1 - 4\!\left(\frac{m_f}{m_S}\right)^2\right]^{3/2}, 
& ff \\[10pt]
\displaystyle
\frac{1}{4}\sqrt{1 - 4\!\left(\frac{m_V}{m_S}\right)^2}\times F(m_V,\,m_S)\,,  & \mathcal{V}\mathcal{V}\,,
\end{cases}
\end{align}
where, $F(m_\mathcal{V},\,m_S)=1-4\,(m_\mathcal{V}/m_S)^2+12\,(m_\mathcal{V}/m_S)^4$ and $\mathcal{V}\in\!Z,\,W^\pm$. Due to non-minimal interaction, the DM can be produced from the bath with a reaction rate that we parametrize as, $\gamma_S\!\equiv\!\xi^4\,T^8/M_P^4$. Using Eq.~\eqref{eq:Y0} we obtain the upper bound: 
\bea
|\xi|\lesssim\! 1.3\,\left(1\,\text{TeV}/m_{\rm{DM}}\right)^{1/4}\,\left(10^{15}\,\text{GeV}/\Trh\right)^{3/4},
\label{SDM1}
\eea
to avoid DM over-abundance.
\\

\noindent
{\textbf{Results and discussions}--} 
To compute the $\gamma$-ray flux in our scenarios, we adopt the Navarro–Frenk–White (NFW) density profile for the Milky Way halo~\cite{Navarro:1996gj,Navarro:2003ew,1989A&A...223...89E,Graham:2005xx,Burkert1995}, 
\begin{equation}
\rho_{\rm DM}(R_{\rm GC})=\frac{\rho_{\rm C}}{(R_{\rm GC}/R_{\rm C})(1+R_{\rm GC}/R_{\rm C})^2},
\end{equation}
with $R_{\rm C}=11~\mathrm{kpc}$ and $\rho_{\rm C}$ is fixed by the local density $\rho_\odot=0.43~\mathrm{GeV\,cm^{-3}}$. 
The differential flux from DM decay is
\begin{equation}
\phi_\gamma(E_\gamma)=\frac{\mathcal{D}}{4\pi\,m_{\rm DM}\tau_{\rm DM}}\,\frac{{\rm d}N}{{\rm d}E_\gamma},
\end{equation}
where
\begin{equation}
\mathcal{D}=\frac{1}{\Delta\Omega}\int_{\Delta\Omega}{\rm d}\Omega\int_0^{s_{\rm max}}{\rm d}s\,\rho_{\rm DM}(s,b,l),
\end{equation}
where, 
$r=\sqrt{s^2+R_\odot^2+2sR_\odot\cos b\cos l}$ with $R_\odot=8.3~\mathrm{kpc}$. For the inner Galactic plane, we obtain $\mathcal{D}\simeq 3\times10^{19}~\mathrm{GeV\,cm^{-2}}$. The photon spectrum ${\rm d}N/{\rm d}E_\gamma$ is evaluated using \texttt{HDMSpectra}~\cite{Bauer:2020jay}. For dark photon–photon oscillations in Eq.~\eqref{eq:oscillation}, corresponding flux is, $\Phi_X=\mathcal{D}/(4\pi m_X\tau_U)$.

Constraints derived from the combined LHAASO and Fermi-LAT observations are presented in Fig.~\ref{fig:plt1}, shown via the blue curves in the coupling–mass plane for different DM decay channels. These bounds apply to feebly interacting DM with masses $\gtrsim$ 1~TeV. In each case, we indicate the region of parameter space where $\tau_{\rm{DM}} < \tau_U$. The vertical dashed lines mark the kinematic thresholds for decay into different final states in each case. For dark photon, RHN and pNGB, we additionally highlight (in light green) the values of $\Trh$ that yield the correct relic abundance for $\mdm \geq 1$ TeV. Achieving the correct abundance for heavier DM requires a lower reheating temperature [cf.Eq.~\eqref{eq:Y0}]. Consequently, the LHAASO+Fermi-LAT constraints are relevant for couplings corresponding to $\Trh \lesssim \{9.5\times 10^{15},\,4.2\times 10^{15},\,3.2\times 10^{15}\}$~GeV for the dark photon, RHN and pNGB, respectively. Both the kinetic mixing parameter and the RHN Yukawa coupling are pushed down to values as small as $\mathcal{O}(10^{-26})$, for masses  $\mathcal{O}(1\,\text{PeV})$, depending on the decay channel. For the same DM mass, bounds on massive pNGB scenario requires $C_{ii}/f_\varphi \lesssim 10^{-31}$   {\color{black} (GeV$^{-1}$)} ($i$ stands for different final states, following Eq.~\eqref{eq:pngb-decay}) for $ZZ$ final state. For the non-minimally coupled scenario, we obtain $\xi \lesssim \mathcal{O}(10^{-14})$ for DM mass $\gtrsim 1$ PeV, considering $ZZ,\,hh$ final states. The cyan shaded region in the bottom right panel, derived from Eq.~\eqref{SDM1}, is disallowed because of super-Planckian $\Trh$. For oscillation-driven photon-dark photon conversion we show bounds in Fig.~\ref{fig:osc}, where LHAASO+Fermi-LAT data provide stringent limit, yielding $\varepsilon \lesssim 10^{-3}$ for $m_X \gtrsim 10$ GeV compared to those obtained from low energy scattering, high energy collider and fixed–target experiments~\cite{Fabbrichesi:2020wbt,Lanfranchi:2020crw}. Interestingly, high luminosity LHC (HL-LHC)~\cite{Curtin:2014cca} can potentially test part of this parameter space, as shown by the red dashed curve. 

{\color{black} We compare our bounds with the latest constraints on DM lifetime derived in~\cite{Boehm:2025qro} using LHAASO data in a model-independent framework, assuming DM decays into two-body SM final states. These are shown in the top left and bottom right panel, corresponding to DM decay into $b\bar{b}$ final states. Note that, these constraints appear stronger than those obtained in our work. This is primarily due to the subtraction of the gamma-ray background arising from cosmic ray interactions. Although such background subtraction can lead to tighter constraints, it also introduces significant systematic uncertainties (as shown by the red band) associated with the modeling of cosmic ray interactions and diffuse gamma-ray emission. Since our primary objective is to place conservative bounds on the DM model parameters, we do not incorporate the effects of the cosmic ray induced gamma-ray background.
}

{\color{black} Before concluding this section, let us see whether the same interactions responsible for DM decay can also generate the observed DM abundance in the present framework.  For the case of dark photon DM decaying through the kinetic-mixing portal, the corresponding inverse decay processes, e.g., $f\bar{f}\to X_\mu$, can populate the dark photon abundance in the early Universe. However, requiring the correct relic abundance constrains the kinetic-mixing parameter to be extremely small, typically $\varepsilon \lesssim \mathcal{O}(10^{-12})$, for dark photon masses in the range $ m_X \in \left[10^{-2},\,10^2\right]~{\rm GeV}$. As evident from the top-left panel of Fig.~\ref{fig:plt1}, such values of $\varepsilon$ are incompatible with the lifetime bounds required for viable DM. A qualitatively similar conclusion also applies to the singlet Majorana DM scenario. On the other hand, dark photons and ALPs can naturally obtain the observed relic abundance through the misalignment mechanism~\cite{Preskill:1982cy,Abbott:1982af,Nelson:2011sf}. In particular, for dark photons this mechanism can successfully account for the entire dark matter abundance over a wide mass range $m_X \in \left[10^{-17},\,10^{-5}\right]~{\rm eV}$, while simultaneously remaining consistent with the DM lifetime constraints (see, {\it e.g.},~\cite{Arias:2012az,Caputo:2021eaa}).
\\
}

\noindent
{\textbf{Conclusions}--We have shown that diffuse Galactic high-energy $\gamma$-ray observations from LHAASO, together with Fermi-LAT, can place meaningful constraints on the mass and couplings of decaying DM produced purely through gravitational interactions. We specifically focused on the diffuse emission from the inner Galactic plane. Additionally, we also looked into the constraints from the outer Galaxy and found to be similar to those from the inner Galaxy. It is worth noting that the constraints derived here depend mainly on the DM decay channels, while their (irreducible) gravitational origin provides insight into the maximum temperature at which the DM abundance was set, offering a window into early Universe cosmology. For the non-minimal coupling scenario, however, both the production as well as the decay rates are fixed by the gravitational interaction. Finally, it is worth mentioning that, for computing the constraints, we have ignored the contribution of conventional astrophysical sources such as Supernova Remnants and Pulsars to the Galactic diffuse $\gamma$-ray flux. Including these conventional sources will result in lower $\gamma$-ray flux from DM decay, thereby leading to even stronger constraints than those presented in this work.
\\

\noindent
{\textbf{Acknowledgments}--BB would like to acknowledge fruitful discussions with Sudhakantha Girmohanta. The work of PS is supported by the National Natural Science Foundation of China (Grant No. 12494574), the National Key R$\&$D Program of China (2021YFA0718500), the Chinese Academy of Sciences (Grant No. E329A6M1) and China's Space Origins Exploration Program.}
\begin{widetext}
\appendix
\section{Photon $\longleftrightarrow$ dark photon conversion probability}
\label{sec:osc}
To compute the oscillation probability, we first perform a field redefinition that brings the kinetic terms into canonical form:
\begin{align}
\mathcal{A}^\mu \equiv
\begin{pmatrix}
A_a^\mu \\[4pt] A_s^\mu
\end{pmatrix}
=
\begin{pmatrix}
1 & 0\\[4pt]
-\varepsilon & 1
\end{pmatrix}
\begin{pmatrix}
A^\mu \\[4pt] X^\mu
\end{pmatrix},
\end{align}
where \(A_a^\mu\) and \(A_s^\mu\) denote the active (photon) and sterile (dark photon) states, respectively. In the canonical basis, the Lagrangian becomes
\begin{align}
\mathcal{L} \supset 
-\frac{1}{4}F_{a\mu\nu}F_a^{\mu\nu}
-\frac{1}{4}F_{s\mu\nu}F_s^{\mu\nu}
+\frac{1}{2}\,\mathcal{A}_\mu^{T}\,\mathbb{M}^2\,\mathcal{A}^\mu
+ \mathcal{O}(\varepsilon^2),
\end{align}
with the mass-squared matrix
\begin{align}
\mathbb{M}^2 =
\begin{pmatrix}
m_{\rm eff}^2 & \varepsilon\, m_X^2 \\[5pt]
\varepsilon\, m_X^2 & m_X^2
\end{pmatrix}\,.
\end{align}
Here, the (1,1) entry represents the effective mass of the photon induced by the medium, which vanishes for propagation in vacuum. To evaluate active--sterile oscillations, we follow the time evolution of the system by solving the corresponding Klein--Gordon equation:
\begin{align}
\left(\omega^2 - k^2 - \mathbb{M}^2 \right)\,
\widetilde{\mathcal{A}}^\mu(\omega, k) = 0,
\end{align}
where \(\widetilde{\mathcal{A}}^\mu(\omega, k)\) denotes the field in Fourier space. In the relativistic limit, $\omega \simeq k \gg m_X,\, m_{\rm eff}$, the system reduces to a linearized Schr\"odinger-like equation~\cite{PhysRevD.37.1237},
\begin{align}
i\,\partial_z\,\mathcal{A}^\mu = \mathcal{H}\,\mathcal{A}^\mu\,,
\end{align}
with the Hamiltonian
\begin{align}
\mathcal{H} =
\begin{pmatrix}
\omega + \Delta_{\rm pl} & \varepsilon\,\Delta_X \\[5pt]
\varepsilon\,\Delta_X & \omega + \varepsilon\,\Delta_X
\end{pmatrix},
\end{align}
where
\begin{align}
\Delta_{\rm pl} \equiv -\frac{m_{\rm eff}^2}{2\omega}, \qquad
\Delta_X \equiv -\frac{m_X^2}{2\omega}\,.
\end{align}
To diagonalize the Hamiltonian, we introduce the rotation matrix
\begin{align}
\mathcal{U} =
\begin{pmatrix}
\cos\theta & -\sin\theta \\
\sin\theta & \cos\theta
\end{pmatrix},
\end{align}
such that $\mathcal{H}_{\rm diag} = \mathcal{U}^\dagger \mathcal{H}\,\mathcal{U}$, with the mixing angle: $\theta = \frac{1}{2}\,\arctan\!\left[\frac{2\varepsilon\,\Delta_X}{\Delta_{\rm pl}-\Delta_X}\right]$. In this basis, the solution takes the form,
\begin{align}\label{eq:A-sol}
A_k^{\rm diag}(z) 
= \exp\!\left[-i\!\int_{z_0}^{z} dz'\,\Theta_k(z')\right] A_k(z_0)\,,
\end{align}
where the eigenvalues of $\mathcal{H}$ are
\begin{align}
\lambda_k = \frac{1}{2}\Big(2\omega + \Delta_{\rm pl} + \Delta_X 
\pm \sqrt{4\varepsilon^2\Delta_X^2 + (\Delta_{\rm pl}-\Delta_X)^2}\Big)\,.
\end{align}
The $z$-dependence in these quantities arises from $m_{\rm eff}$, which may be treated as approximately constant whenever it varies on scales much larger than~$k^{-1}$~\cite{Brahma:2023zcw,Ismail:2024bjw}. Assuming the initial state at $z_0=0$ is a pure dark photon, the probability of detecting a SM photon after traveling a distance $z$ is
\begin{align}
P_{X\to\gamma}
= \left|\langle A_a(z)\,|\,A_s(0)\rangle\right|^2
= \sin^2(2\theta)\,
\sin^2\!\left(\frac{z\,\Delta\lambda}{2}\right)\simeq 
\frac{4\varepsilon^2\,\Delta_X^2}{(\Delta_X - \Delta_{\rm pl})^2}
\sin^2\!\left[\frac{z}{2}\,(\Delta_X-\Delta_{\rm pl})\right]
+ \mathcal{O}(\varepsilon^3)\,,
\label{eq:prob}
\end{align}
where,
\begin{align}
\Delta\lambda = 
\sqrt{4\varepsilon^2\Delta_X^2 + (\Delta_{\rm pl}-\Delta_X)^2}\,.
\end{align}
At resonance, where $\Delta_{\rm pl} = \Delta_X$, the conversion becomes maximal: $P^{\rm res}_{X\to\gamma}
= \sin^2\!\left(\frac{z\,\varepsilon\,m_X^2}{2\omega}\right)$. In the interstellar medium the plasma frequency is $\omega_{\rm pl}\sim10^{-12}\,\mathrm{eV}$, implying an effective photon mass $m_{\rm eff}=\omega_{\rm pl}\ll m_X$. Under this condition, Eq.~\eqref{eq:prob} reduces to 
\(P_{X\to\gamma}=4\varepsilon^2\sin^2\!\left(\tfrac{z m_X^2}{4\omega}\right)\). For Galactic propagation distances, the oscillatory term averages to $1/2$, yielding a constant conversion probability: $P_{X\to\gamma}=2\varepsilon^2$.
\section{pNGB EFT}
\label{sec:pngb-int}
Above the electoweak symmetry breaking (EWSB) scale, the pNGB Lagrangian reads,
\begin{align}
& \mathcal{L}_\varphi\supset\frac{1}{2}\left[(\partial\varphi)^2 - m_\varphi^2\,\varphi^2\right]
+ \left(\frac{g_1}{4\pi}\right)^2 C_{BB}\, \frac{\varphi}{f_\varphi} B_{\mu\nu}\widetilde{B}^{\mu\nu}
+ \left(\frac{g_2}{4\pi}\right)^2 C_{WW}\, \frac{\varphi}{f_\varphi} W^i_{\mu\nu}\widetilde{W}_i^{\mu\nu}+C_{GG}\,\frac{\alpha_s}{4\pi}\,\frac{\varphi}{f_\varphi}\,G_{\mu\nu}^a\,\widetilde{G}^{a\,\mu\nu}
\nonumber\\&
+ \left(\frac{g_1}{4\pi}\right)\left(\frac{g_2}{4\pi}\right)
C_{BW}\, \frac{\varphi}{f_\varphi} B_{\mu\nu}\widetilde{W}^{3\,\mu\nu}\,,
\label{eq:EFTLag}
\end{align}
where $W_{\mu\nu}^i$ and $B_{\mu\nu}$ denote the field-strength tensors of the
$\mathrm{SU}(2)_W$ and $\mathrm{U}(1)_Y$ gauge groups before EWSB, and the corresponding dual field strengths are defined by
$\widetilde{W}^{i\,\mu\nu} = \epsilon^{\mu\nu\rho\sigma} W^i_{\rho\sigma}/2$ and
$\widetilde{B}^{\mu\nu} = \epsilon^{\mu\nu\rho\sigma} B_{\rho\sigma}/2$. We point out that pNGB couples with all of the SM fermions as well as with the SM Higgs~\cite{Brivio:2017ije,Bauer:2017ris,Bauer:2020jbp}. Here $f_\varphi$ is the pNGB decay constant that is related to the relevant new-physics scale by $\Lambda=4\pi f_\varphi$ and $C_{ii}$'s are the Wilson coefficients. For simplicity, we are considering interactions only with the gauge bosons. These effective interactions may arise from integrating out heavy fermions charged under the electroweak group at one loop. Consequently, the coefficients are suppressed by loop factors of order $1/(4\pi)^2$. A commonly used choice rescales the parameters
\begin{align}
&\{C_{WW},\, C_{BB},\, C_{BW}\}
\to\left\{
\left(\frac{g_2}{4\pi}\right)^2 C_{WW},\;
\left(\frac{g_1}{4\pi}\right)^2 C_{BB},\;
\frac{g_1 g_2}{(4\pi)^2} C_{BW}
\right\}\,,   
\end{align}
absorbing both gauge couplings and loop factors into the definitions of the Wilson coefficients. We adopt the normalization in Eq.~\eqref{eq:EFTLag} throughout our analysis. Below the EWSB scale, interactions with the gauge boson mass eigenstates take the form
\begin{align}
&\mathcal{L}_\varphi\supset\frac{1}{2}(\partial\varphi)^2 
- \frac{1}{2} m_\varphi^2\,\varphi^2
+ \frac{\alpha}{4\pi}\frac{\varphi}{f_\varphi} \Big(
C_{\gamma\gamma} F\widetilde{F}+2\,C_{\gamma Z} F\widetilde{Z}
+ C_{ZZ}\,Z\widetilde{Z}
+ \frac{2}{s_W^2}\,C_{WW}\,W^+ \widetilde{W}^-
\Big),
\label{eq:EWSB_Lag}
\end{align}
where the electroweak field strengths satisfy
$F\widetilde{Z} \equiv \frac{1}{2} \epsilon^{\mu\nu\rho\sigma}
F_{\mu\nu} Z_{\rho\sigma}$ and $s(c)_W \equiv \sin(\cos)\theta_W \simeq 0.48 (0.87)$ being the measures of weak mixing angle in the $\overline{\text{MS}}$ scheme~\cite{PDG2024}.
The coefficient $C_{WW}$ remains unchanged, while the remaining three couplings are,
\begin{align}
C_{\gamma\gamma} &\equiv C_{BB} + C_{WW} + C_{BW}, \\
C_{\gamma Z} &\equiv C_{WW}\frac{c_W}{s_W}
- C_{BB}\frac{s_W}{c_W}
+ \frac{1}{2} C_{BW}\left( \frac{c_W}{s_W} - \frac{s_W}{c_W} \right), \\
C_{ZZ} &\equiv C_{WW}\frac{c_W^2}{s_W^2}
+ C_{BB}\frac{s_W^2}{c_W^2}
- C_{BW}\,.
\end{align}
\end{widetext}
\vspace{-0.398in}
\bibliographystyle{utphys}
\bibliography{bibliography}

\end{document}